\documentclass[journal, twoside]{IEEEtran}
\usepackage[latin1]{inputenc}
\usepackage{color}
\usepackage[tight]{subfigure}
\usepackage[table]{xcolor}
\usepackage{colortbl}
\usepackage{cite}

\newcommand{\equ}[1]{{$($\ref{#1}$)$}}
\newcommand{\fig}[1]{{Figure \ref{#1}}}

\ifCLASSINFOpdf
\else
  \usepackage[dvips]{graphicx}
\fi
\begin{document}
%
\title{Transpolarising Trihedral Corner Reflector Characterisation Using a GB-SAR System}
%
%
%

\author{Pere~J.~Ferrer,
        Carlos~López-Martínez,
        Albert~Aguasca,
        Luca~Pipia,
        José~M.~González-Arbesú,
        Xavier~Fabregas,
        and~Jordi~Romeu
\thanks{Manuscript Received Sep XX, 2010; revised January XX, 2011. This work was partially supported by the Spanish CICYT projects TEC2007-66698-C04-01/TCM, TEC2007-65690, TEC2008-06764-C02-01, TEC2009-13897-C03-01, and CONSOLIDER CSD2008-00068, and by the Ramón y Cajal Programme.
}
\thanks{P.J. Ferrer, J.M. González-Arbesú, and J. Romeu are with AntennaLab group, and C. López-Martínez, A. Aguasca and X. Fabregas are with RSLab group, both in the Department of Signal Theory and Communications, Universitat Politecnica de Catalunya, 08034, Barcelona, Spain. e-mail: pj.ferrer@tsc.upc.edu, carlos.lopez@tsc.upc.edu.}
\thanks{L. Pipia is with Institut Cartogràfic de Catalunya (ICC), Barcelona, Spain. e-mail: luca.pipia@icc.cat.}
\thanks{Color versions of one or more of the figures in this paper are available online
at http://ieeexplore.ieee.org.}
\thanks{Digital Object Identifier.}}

\maketitle

\begin{abstract}
The use of a low profile, light weight and easy to fabricate transpolarising surface placed on one side of a trihedral corner reflector (TCR) as polarimetric calibrator is presented in this letter. The transpolarising-TCR (TTCR) presents a high backscattered cross-polar response contrary to standard TCRs. The performance of this device has been tested at X-band using the UPC GB-SAR.
\end{abstract}


%
\IEEEpeerreviewmaketitle

%
%
%
%

\section{Introduction}

\IEEEPARstart{P}{olarimetric} SAR (PolSAR) systems have caught the interest of the research community since they are able to provide much more information than conventional single polarisation systems. This interest will gradually increase since several spaceborne PolSAR missions have been launched in the recent years: ALOS-PALSAR, RADARSAT-2 and TERRASAR-X. Thus, reliable polarimetric calibration procedures and techniques are mandatory. 

PolSAR calibration is commonly performed by using dihedral corner reflectors tilted 45$^\circ$, although they have a narrow beam-width response in elevation. Trihedral corner reflectors are well known SAR data calibrators, as they provide a high backscattering RCS response for a wide range of incident angles. Nevertheless, the TCRs lack of a cross-polar response, making them limited for full polarimetric calibration, as stated from the scattering matrix of a TCR \equ{eq:transpol_rcs_tcr},
\begin{equation}
\emph{\textbf{S}}_{TCR} = \left[ \begin{array}{c c} S_{HH} & S_{HV}\\ S_{VH} & S_{VV} \end{array} \right] = \emph{A}\left[ \begin{array}{c c} 1 & 0\\ 0 & 1 \end{array} \right],
\label{eq:transpol_rcs_tcr}
\end{equation}

\noindent where $A\;=\;S_0e^{j\phi_0}$ is a constant, and $H$ and $V$ stand for horizontal and vertical polarisation, respectively. Therefore, a TCR providing cross-polar response, that is, a transpolarising trihedral corner reflector (TTCR), would be characterised by a scattering matrix \equ{eq:transpol_rcs_ttcr} with anti-diagonal non-zero values,
\begin{equation}
\emph{\textbf{S}}_{TTCR} = \left[ \begin{array}{c c} S_{HH} & S_{HV}\\ S_{VH} & S_{VV} \end{array} \right] = \emph{A}\left[ \begin{array}{c c} 0 & 1\\ 1 & 0 \end{array} \right].
\label{eq:transpol_rcs_ttcr}
\end{equation}

Different solutions appeared in the 90's to realise such TTCRs, taking advantage of transpolarising surfaces composed of grids, fins or corrugations \cite{michelson1990_1, sheen1992_1, macikunas1993_1, michelson1995_1}. TTCRs are passive devices, but they are frequency dependent due to the design of the periodic inclusions. Moreover, a TTCR could even have non-cross-polarising applications, like a gridded TCR used as polarisation selective calibrator in order to produce scattering matrices with only horizontal ($S_{HH}$) or vertical ($S_{VV}$) channels by properly choosing the azimuth angle of incidence \cite{lavalle2009_1}.
The transpolarising surfaces, also referred as twist or depolarising reflectors, are well known polarisation conversion surfaces \cite{lerner1965_1, hanfling1981_1, fusco2007_1}, and they are characterised by a 90$^\circ$ rotation in reflection of the incident electric field polarisation. They are also applied as trans-reflectors to reduce the secondary reflector blockage in Cassegrain antenna designs \cite{hannan1961_1, kastner1982_1, brooker2000_1}. Such transpolarising surfaces are placed on the bottom side of the TCR, with the fins or corrugations aligned at 45$^\circ$ with respect to the incident electric field polarisation, thus producing a TCR with cross-polarisation response. All these TTCR designs present some fabrication issues, like the accuracy of the grid array above the absorbing material, or the heavy and bulky piece of metal required to produce the $\lambda$/4 corrugations. A low profile and light weight transpolarising surface composed of a periodic arrangement of metallic square patches with diagonal slots over a metal ground plane was presented in \cite{ferrer2006_1}. The advantages of this transpolarising surface applied as PolSAR calibrator were discussed in \cite{ferrer2007_igarss_1}.

In this letter, the performance of a TTCR realised with the proposed transpolarising surface is experimentally assessed for PolSAR calibration purposes. A measurement campaign has been carried out at the Campus Nord of the Universitat Politècnica de Catalunya (Barcelona, Spain), using a ground-based SAR system, the so called UPC X-band GB-SAR \cite{aguasca2004_1, pipia2007_1}.

\section{Transpolarising Surface Design}
\begin{figure}[!h]
\centering
\includegraphics[width=7.75cm]{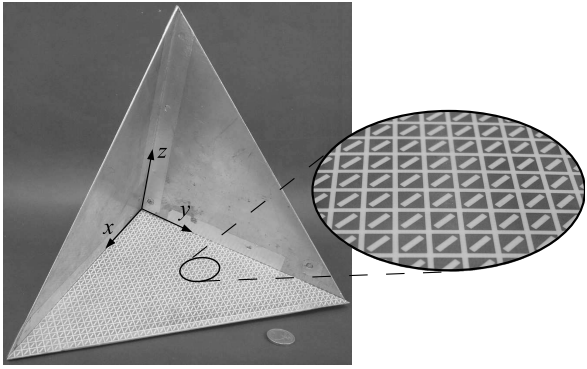}
\caption{Fabricated transpolarising trihedral corner reflector (TTCR), with a detail of the transpolarising surface placed on the bottom side of the TCR. \label{f:transpol_gbsar_fab_ttcr}}
\end{figure}

A transpolarising surface has been designed and fabricated to operate around 9.65 GHz (X-band), according to the design guidelines presented in \cite{ferrer2007_1}. The transpolarising surface is basically composed of square patches with inner diagonal slots etched on a grounded dielectric substrate. The square patches have a width of 4.8 mm, with a gap of 1 mm between adjacent patches. The overall size of the unit cell is then 5.8 mm, which corresponds to $\lambda$/5.4. The diagonal slot has a length of 4.8 mm and a width of 1.4 mm. The prototype has been fabricated using standard photo-etching techniques applied on a 1.52 mm thick Rogers RO4003C ($\varepsilon_r$ = 3.38, tan $\delta$ = 0.0027) dielectric substrate. The overall thickness of the transpolarising surface is 1.52 mm, that is, $\lambda$/20, which is much smaller than the $\lambda$/4 thickness required for the fabrication of the corrugations \cite{michelson1995_1}. The fabricated triangular shaped transpolarising surface is then placed on the bottom side of a TCR forming a TTCR, as shown in \fig{f:transpol_gbsar_fab_ttcr}. 

\begin{figure}[!h]
\centering
\includegraphics[width=7.75cm]{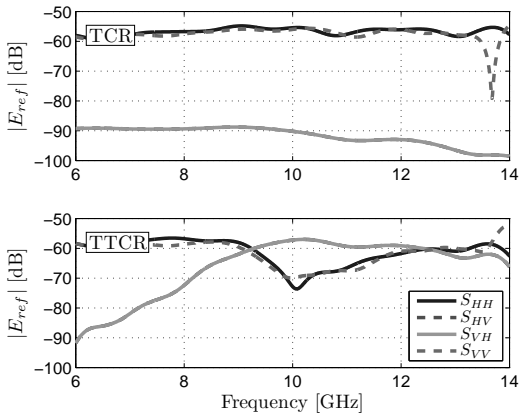}
\caption{Measured reflection response of the TCR and TTCR in the anechoic chamber for $\tau = 0^\circ$. \label{f:transpol_meas_ttcr}}
\end{figure}

The standard TCR and the TTCR have been measured in the UPC anechoic chamber using two broadband ridged horn antennas placed in a bi-static configuration. The TCR is oriented at boresight, 
with a roll angle $\tau$ of 0$^\circ$, an azimuth angle $\phi$ of 45$^\circ$ and an elevation angle $\theta$ of 54.74$^\circ$ \cite{michelson1995_1}, for which the maximum response of a TCR/TTCR is expected, as also confirmed in \cite{gennarelli1998_1}$-$\cite{hanninen2006_1} using different numerical methods. Measured co-polar ($S_{HH}$ and $S_{VV}$) and cross-polar ($S_{HV}$ and $S_{VH}$) responses are plotted in \fig{f:transpol_meas_ttcr}. Note that the cross-polar channels are equal ($S_{HV}$ = $S_{VH}$), for reciprocity of the measuring antenna system. The measured results show that the standard TCR presents a broadband response with a co-polar to cross-polar ratio of about 30 dB. On the other hand, the transpolarising TCR produces a high cross-polar response around 10 GHz with a cross-polar ratio of more than 14 dB, whereas the cross-polar ratio around 9.65 GHz is about 8 dB, which is not maximum due to a slight frequency shift of the transpolarising surface when placed inside the TCR. 

\begin{figure}[!h]
\centering
\includegraphics[width=7.75cm]{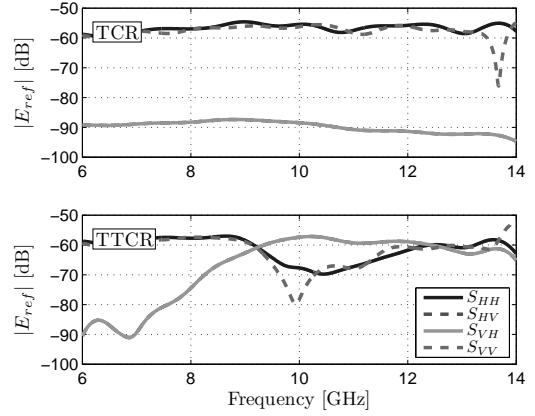}
\caption{Measured reflection response of the TCR and TTCR in the anechoic chamber for $\tau = 90^\circ$. \label{f:transpol_meas_ttcr_90}}
\end{figure}

Moreover, the TTCR presents an angular $\tau$ dependence around its axis due to the geometry of the transpolarising surface. The same performance is obtained at every 90$^\circ$ rotation, contrary to the roll-invariance of a TCR. This fact is evidenced in \fig{f:transpol_meas_ttcr_90} for an angular $\tau$ rotation of 90$^\circ$.

\begin{figure}[!h]
\centering
\includegraphics[width=7.75cm]{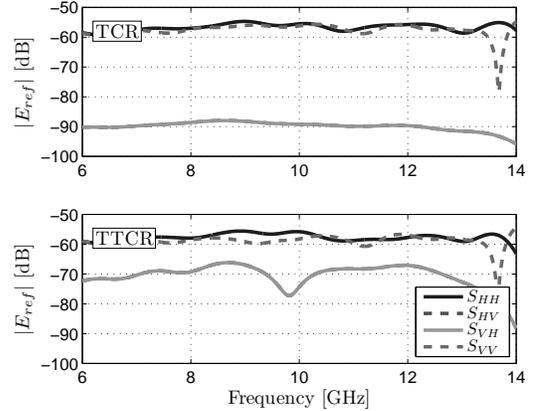}
\caption{Measured reflection response of the TCR and TTCR in the anechoic chamber for $\tau = 45^\circ$. \label{f:transpol_meas_ttcr_45}}
\end{figure}

Another consequence of the angular dependence is that the TTCR behaves almost like a standard TCR for $\tau$ equal to 45$^\circ$, but with higher cross-polar responses, as shown in \fig{f:transpol_meas_ttcr_45}.

The scattering matrices $($\ref{eq:transpol_rcs_tcr}$)-($\ref{eq:transpol_rcs_ttcr}$)$ have been extracted for the TCR and TTCR, considering an orientation $\tau = 0^\circ$. The results show the expected behaviour for the TCR, whereas the TTCR presents higher co-polar components that may be produced by the suboptimal response found at 9.65 GHz. Note that no external calibration has been applied to these results.
\begin{equation}
\begin{array}{c}
\emph{\textbf{S}}_{TCR}^{meas} = \emph{A}\left[ \begin{array}{c c} 1.00_{\angle\;0^\circ} & 0.02_{\angle\;30^\circ}\\ 0.02_{\angle\;30^\circ} & 0.93_{\angle\;13^\circ} \end{array} \right]\\
\\
\emph{\textbf{S}}_{TTCR}^{meas} = \emph{A}\left[ \begin{array}{c c} 0.43_{\angle\;-22^\circ} & 1.00_{\angle\;0^\circ}\\ 1.00_{\angle\;0^\circ} & 0.36_{\angle\;-77^\circ} \end{array} \right]
\end{array}
\label{eq:transpol_rcs_tcr_ttcr_anechoic}
\end{equation}

\begin{figure}[!h]
\centering
\subfigure[][TCR $\theta_1$]{\includegraphics[width=4.25cm]{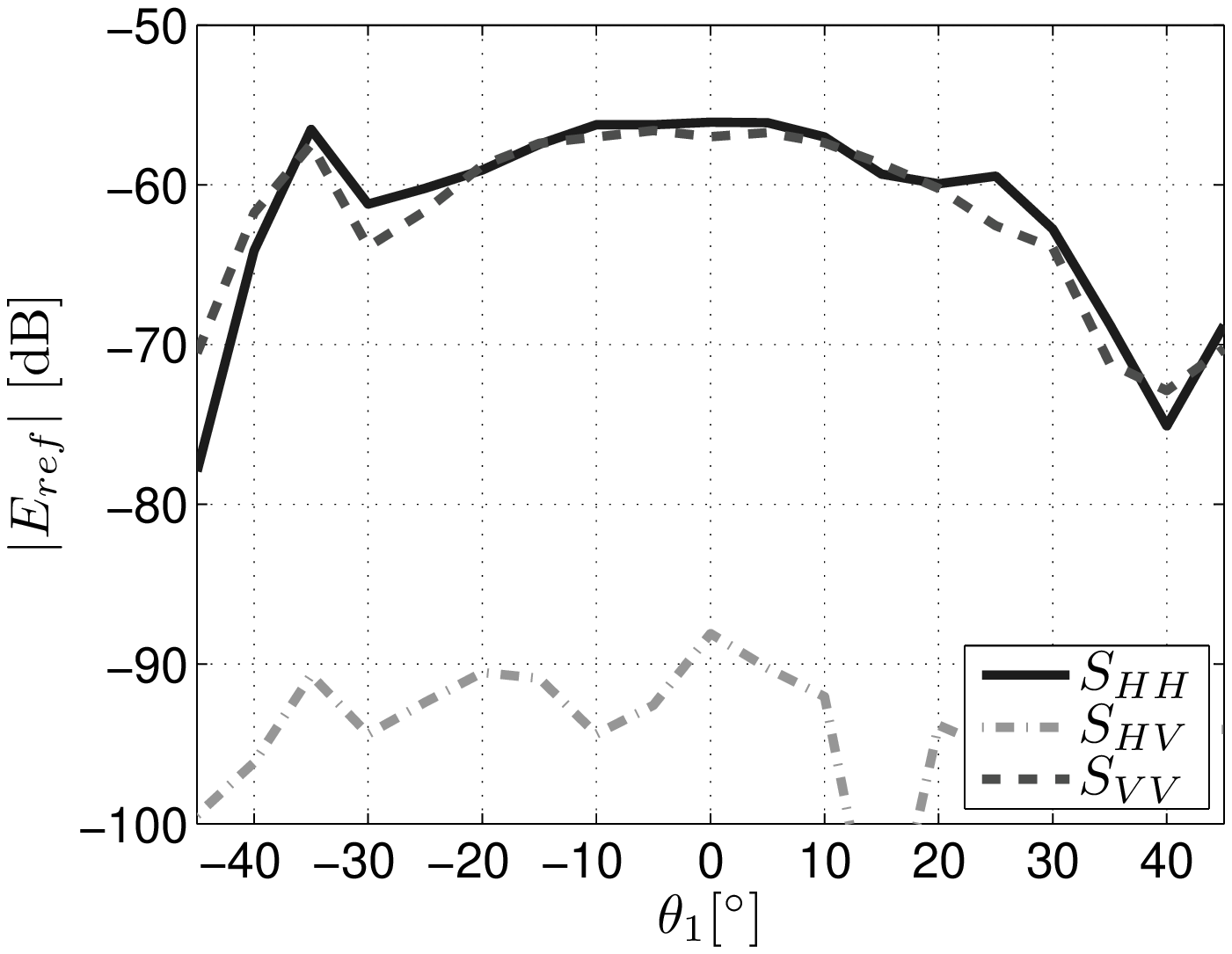}}\quad
\subfigure[][TTCR $\theta_1$]{\includegraphics[width=4.25cm]{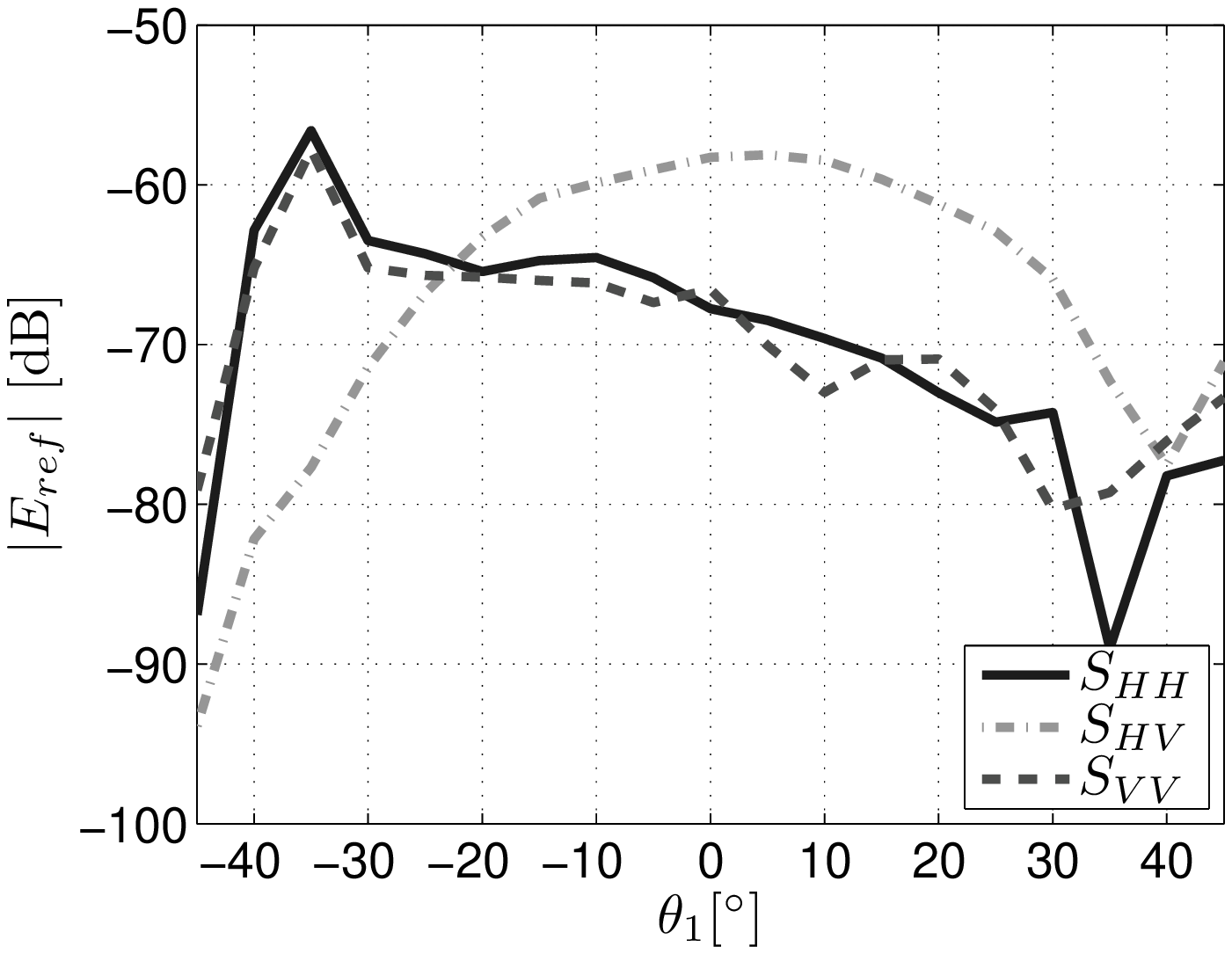}}\quad
\subfigure[][TCR $\phi_1$]{\includegraphics[width=4.25cm]{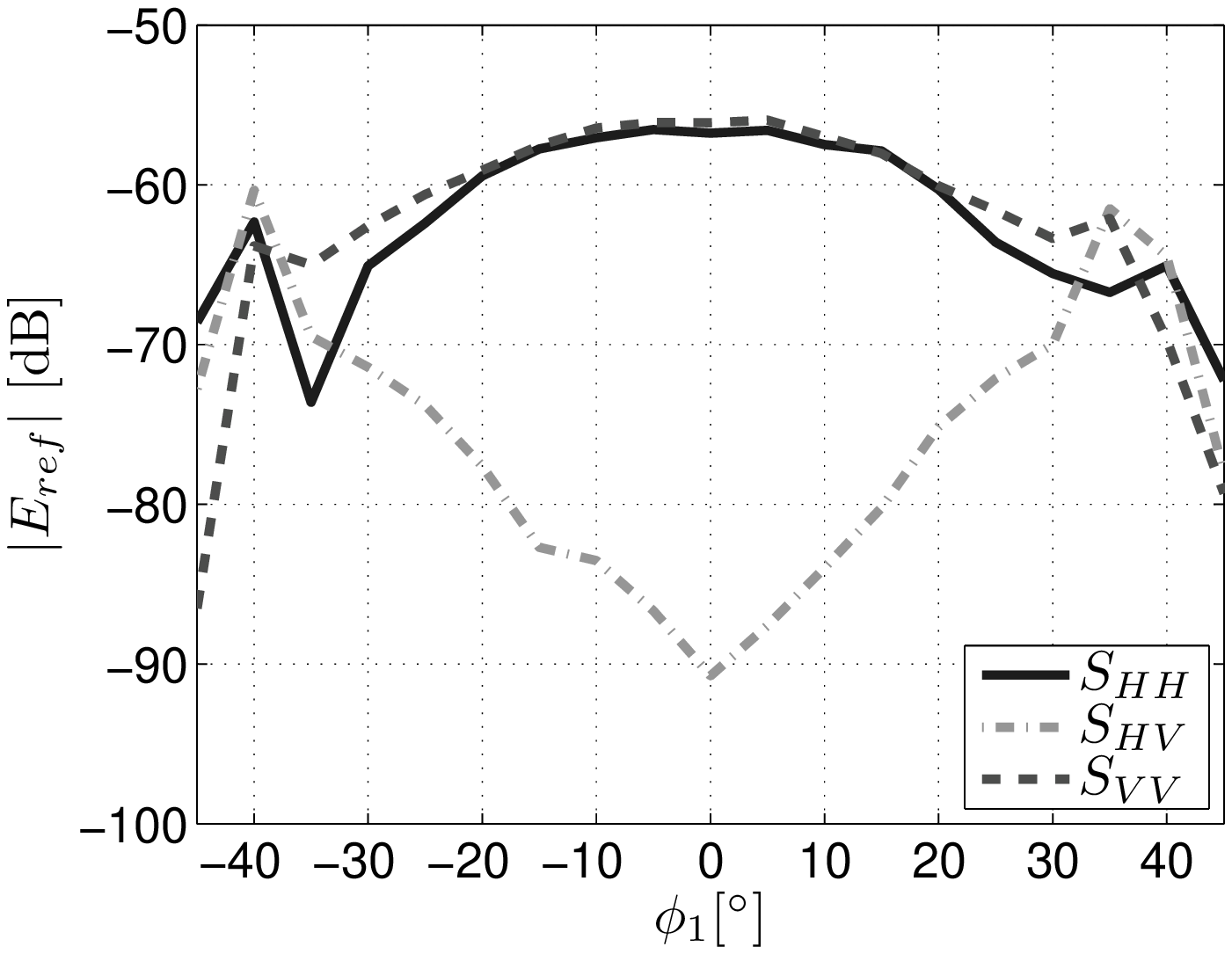}}\quad
\subfigure[][TTCR $\phi_1$]{\includegraphics[width=4.25cm]{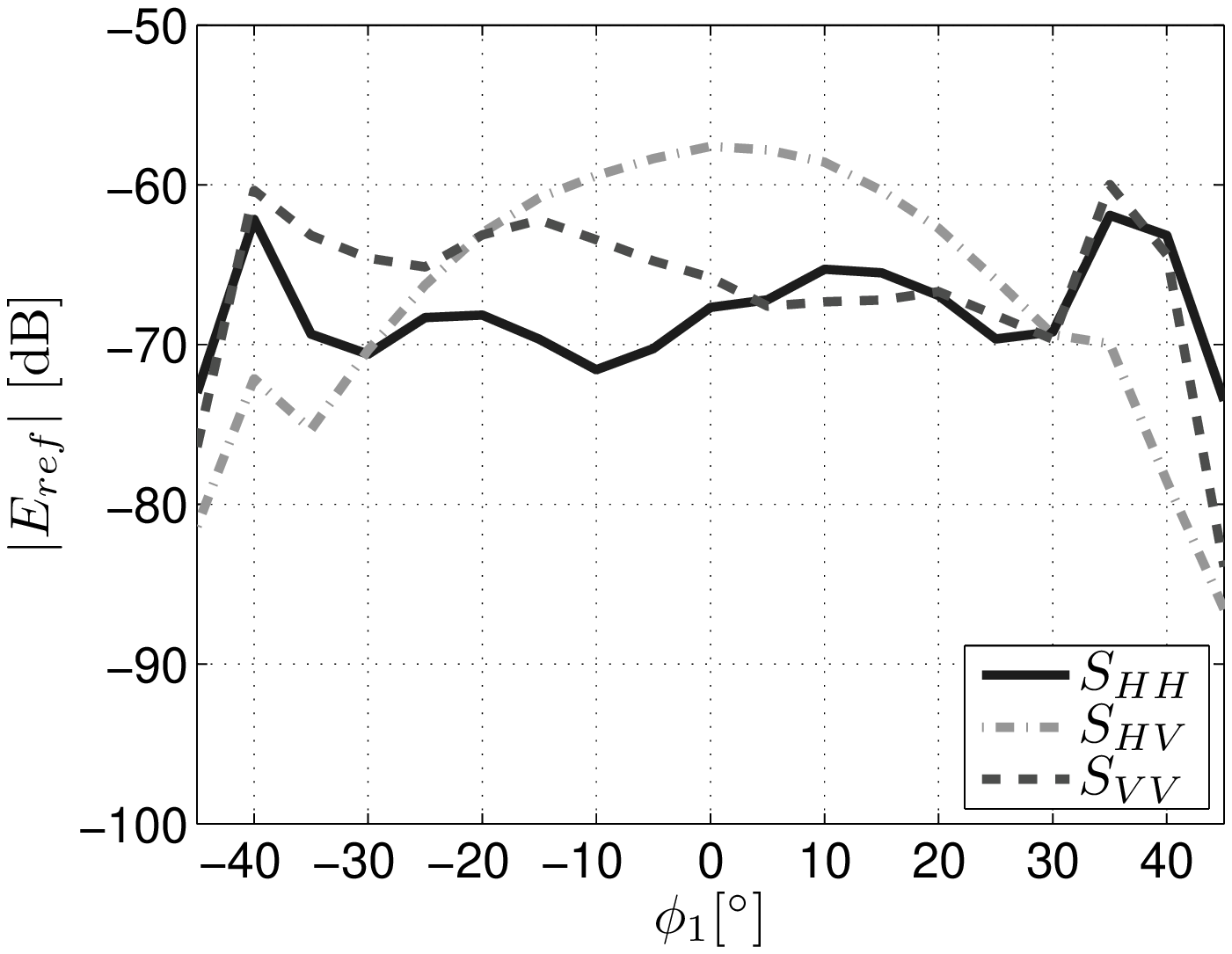}}\quad
\subfigure[][TCR $\tau$]{\includegraphics[width=4.25cm]{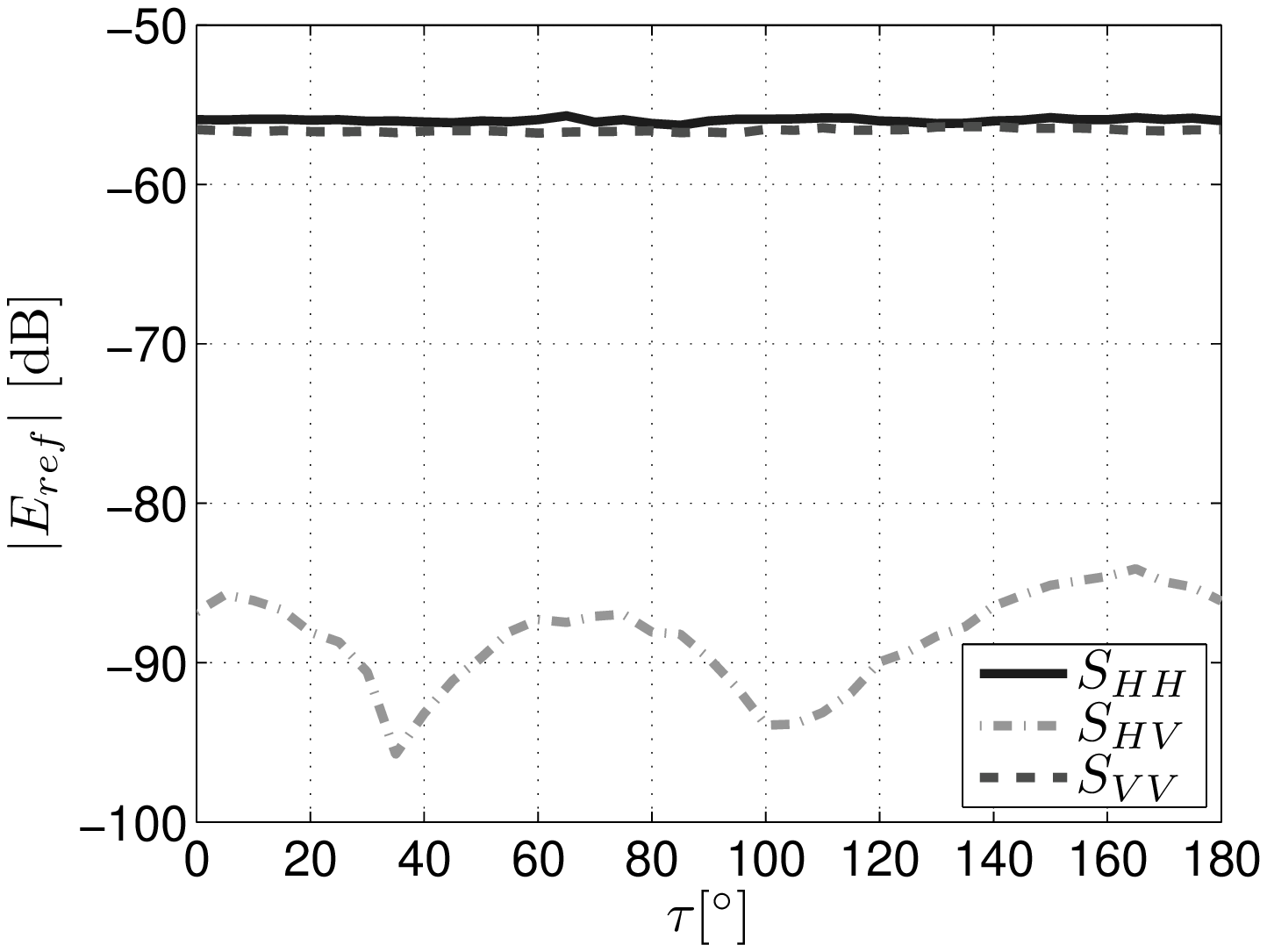}}\quad
\subfigure[][TTCR $\tau$]{\includegraphics[width=4.25cm]{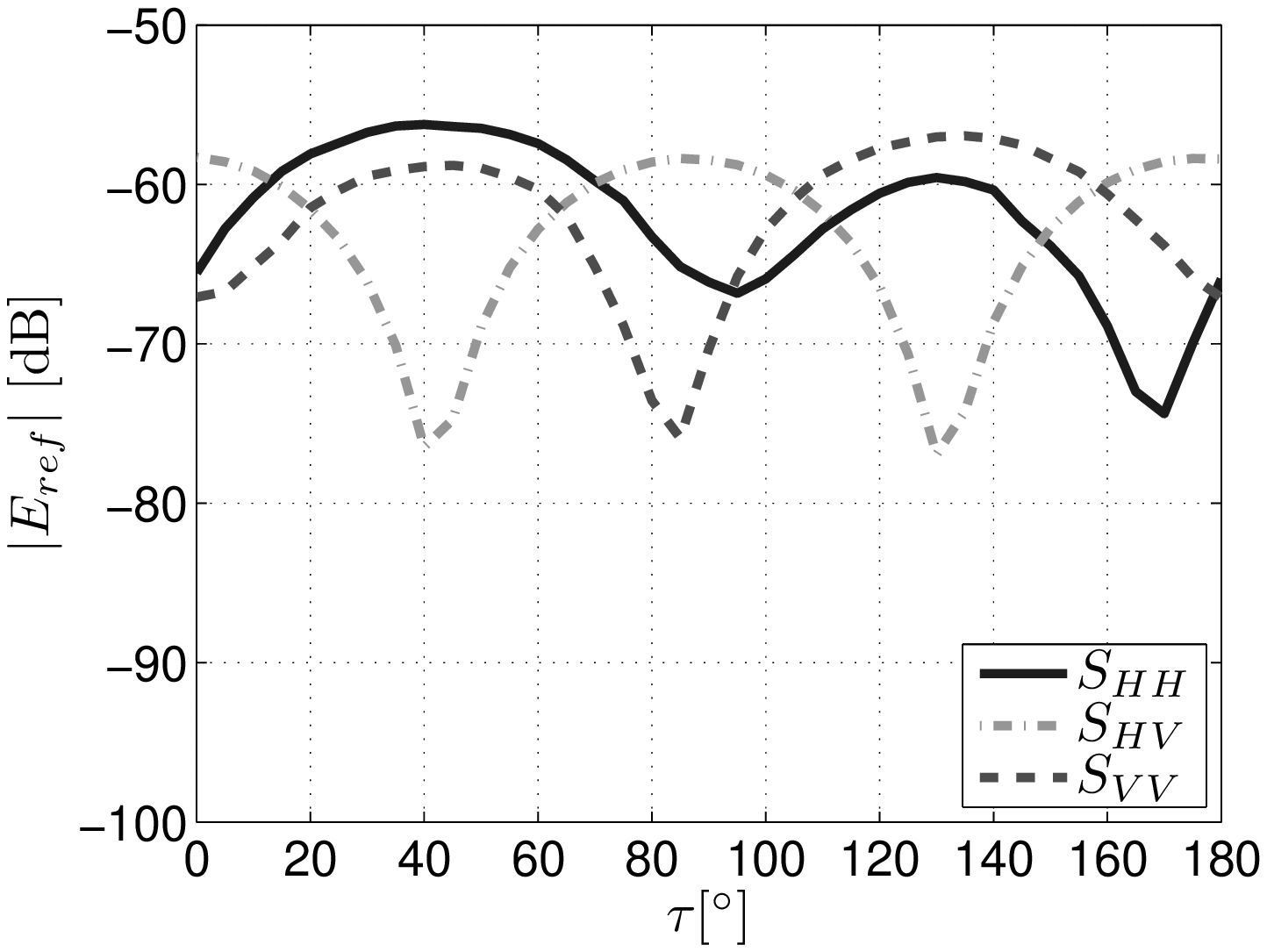}}\quad
\caption{Measured angular response for the TCR/TTCR at 9.65 GHz. \label{f:transpol_meas_tcr_ttcr_angular}}
\end{figure}

The angular response around boresight of the TCR/TTCR has been also studied. In particular, measurements have been carried out at 9.65 GHz for different angles $-45^\circ\leq\theta_1\leq45^\circ$, $-45^\circ\leq\phi_1\leq45^\circ$, and $0^\circ\leq\tau\leq180^\circ$, as presented in \fig{f:transpol_meas_tcr_ttcr_angular}, where $\theta_1 = 54.75^\circ$ and $\phi_1 = 45^\circ$. The angular response of the TCR shows that its maximum response is found at boresight. However, the TTCR slightly improves its response at boresight when $0^\circ \leq \theta_1 \leq 30^\circ$, because the transpolarising surface placed inside the TTCR is illuminated with a higher grazing angle, thus enhancing the transpolarisation effect.

\section{Field Measurement Results}
Full polarimetric field measurements at 9.65 GHz (X-band) have been carried out with a GB-SAR system to assess the performance of the designed TTCR. 

\subsection{Measurement Setup}

\begin{figure}[!h]
\centering
\includegraphics[width=7.75cm]{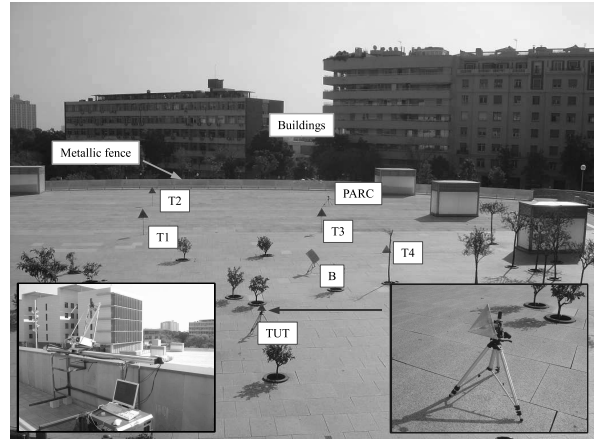}
\caption{Measurement scenario at the Campus Nord UPC. The trihedral under test (TUT) and the GB-SAR system are shown in the inset. \label{f:transpol_gbsar_scenario}}
\end{figure}

The GB-SAR is mounted on the top of a terrace at the Campus Nord UPC facing towards a flat square with some small trees, as shown in \fig{f:transpol_gbsar_scenario}. The trihedral under test (TUT) has been placed in the middle of the measurement scenario, which is also complemented with different reference point scatters: four conventional TCRs (T1-T4), which provide a pure co-polar response; one Bruderhedral (B) \cite{silverstein1997_1}, tilted 45$^\circ$, which provides a passive pure cross-polar response; and a polarimetric active radar calibrator (PARC) \cite{freeman1990_1}, tilted 45$^\circ$, thus producing co-polar and cross-polar responses, making its signature to be present in the results for the four terms of the measured scattering matrix. The measurement parameters of the UPC GB-SAR system can be found in \cite{pipia2007_1} (Table 1).

\subsection{Measured Results}
\begin{figure}[!h]
\centering
\includegraphics[width=7.75cm]{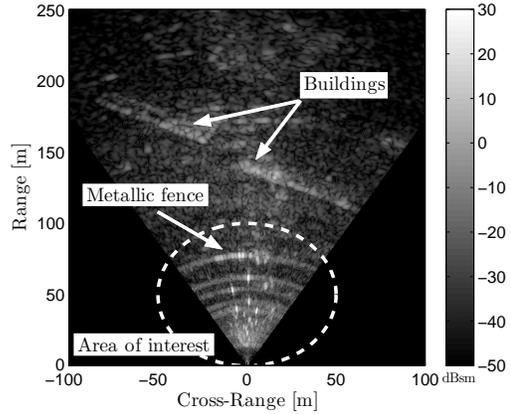}
\caption{Measured $|S_{HH}|$ channel of the scenario at full range.\label{f:transpol_gbsar_meas_scenario}}
\end{figure}

The measured $|S_{HH}|$ channel of the scenario at full range is plotted in \fig{f:transpol_gbsar_meas_scenario}. The maximum range in the measurements is about 250 m, including the metallic fence and the buildings, although the trihedral under test and the reference scatters are located at a range below 70 m.

\begin{figure}[!h]
\centering
\subfigure[][TCR $|S_{HH}|$]{\includegraphics[height=7cm]{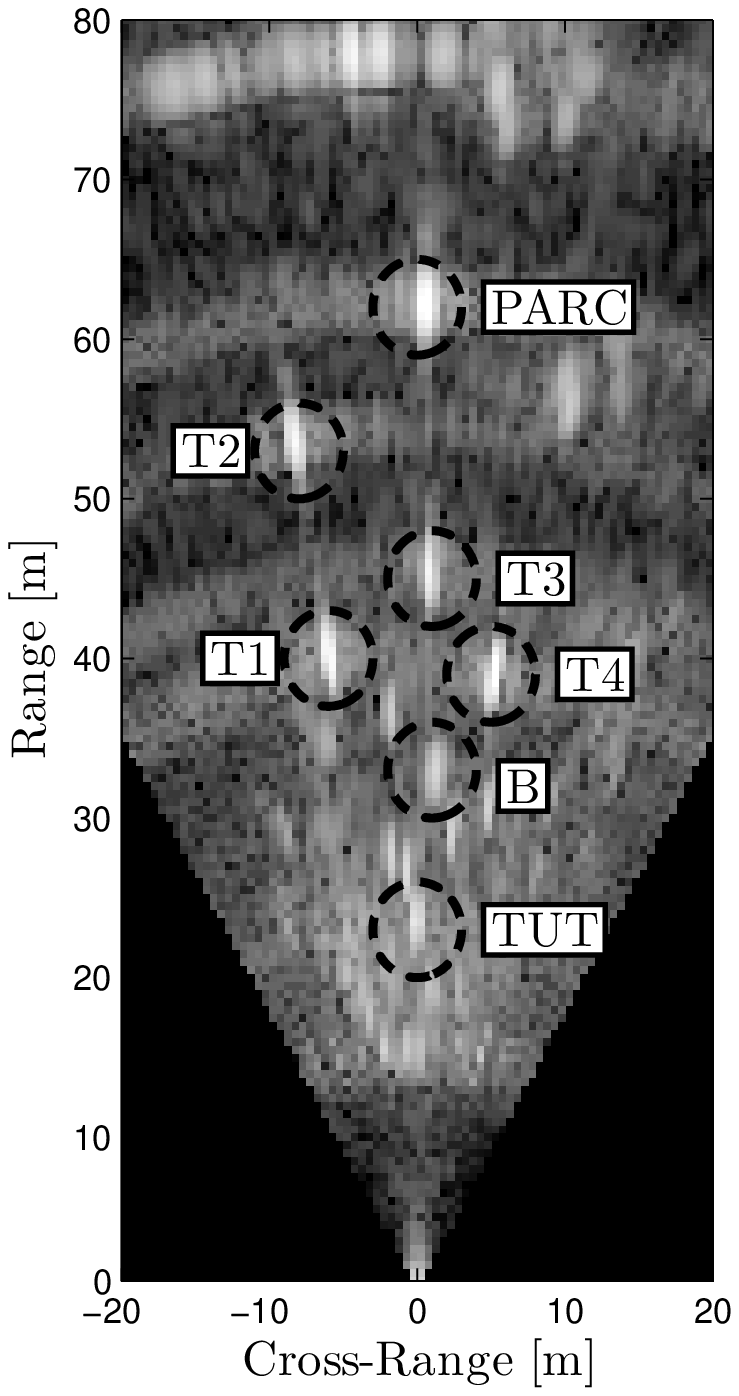}}\quad
\subfigure[][TCR $|S_{HV}|$]{\includegraphics[height=7cm]{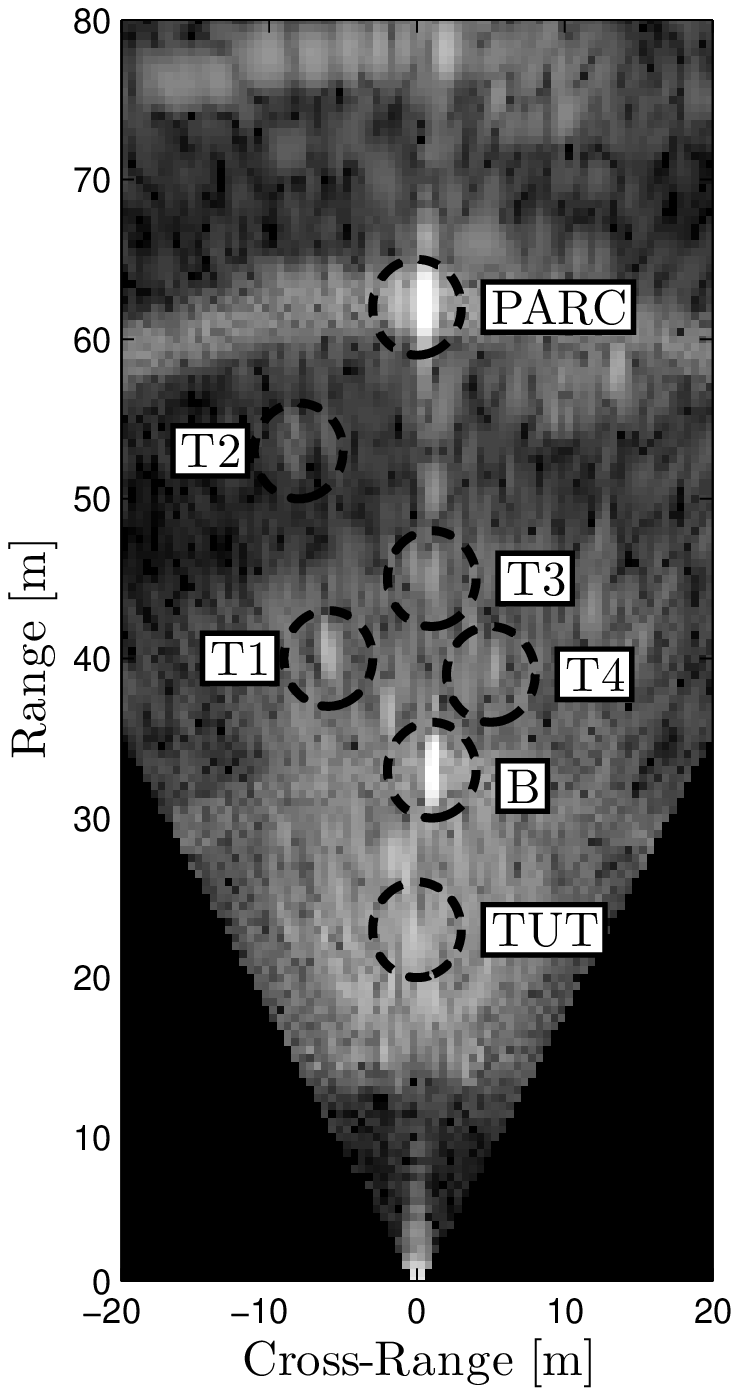}}\quad
\subfigure[][TCR $|S_{VH}|$]{\includegraphics[height=7cm]{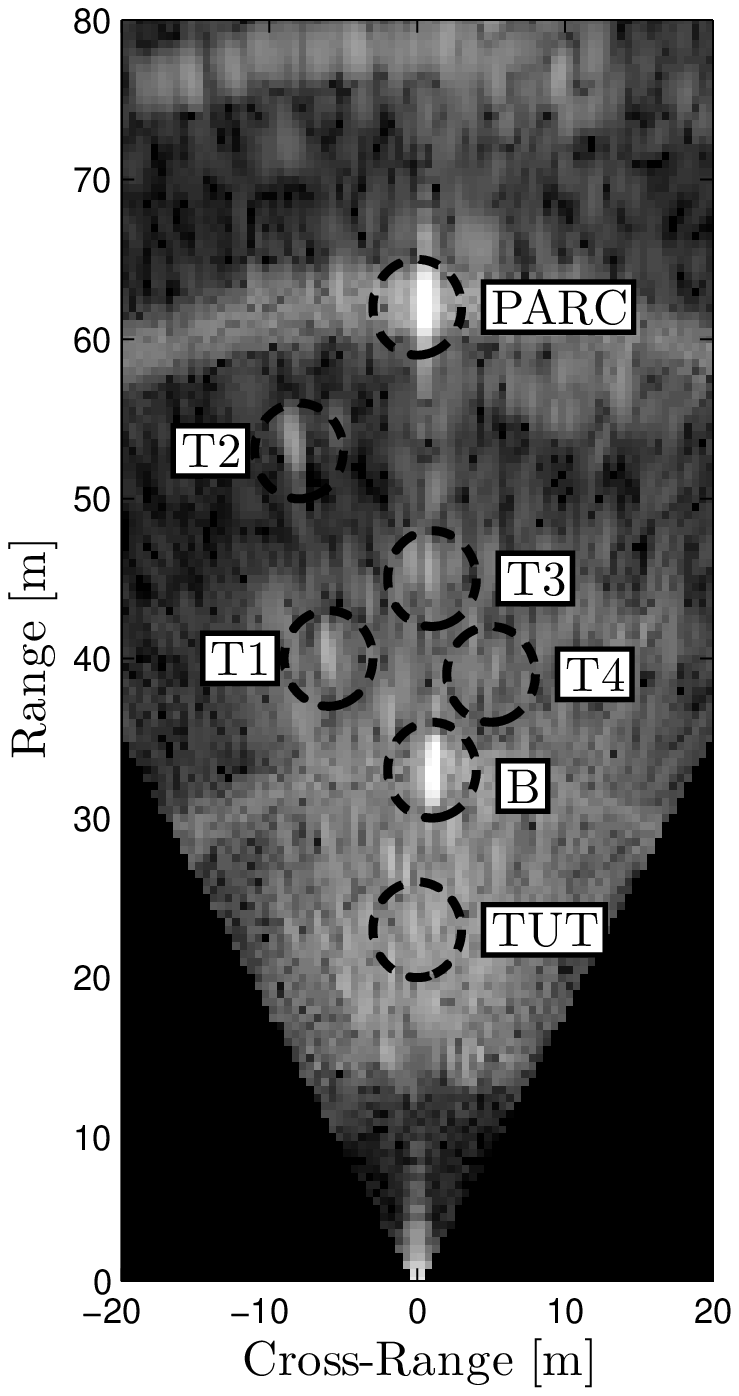}}\quad
\subfigure[][TCR $|S_{VV}|$]{\includegraphics[height=7cm]{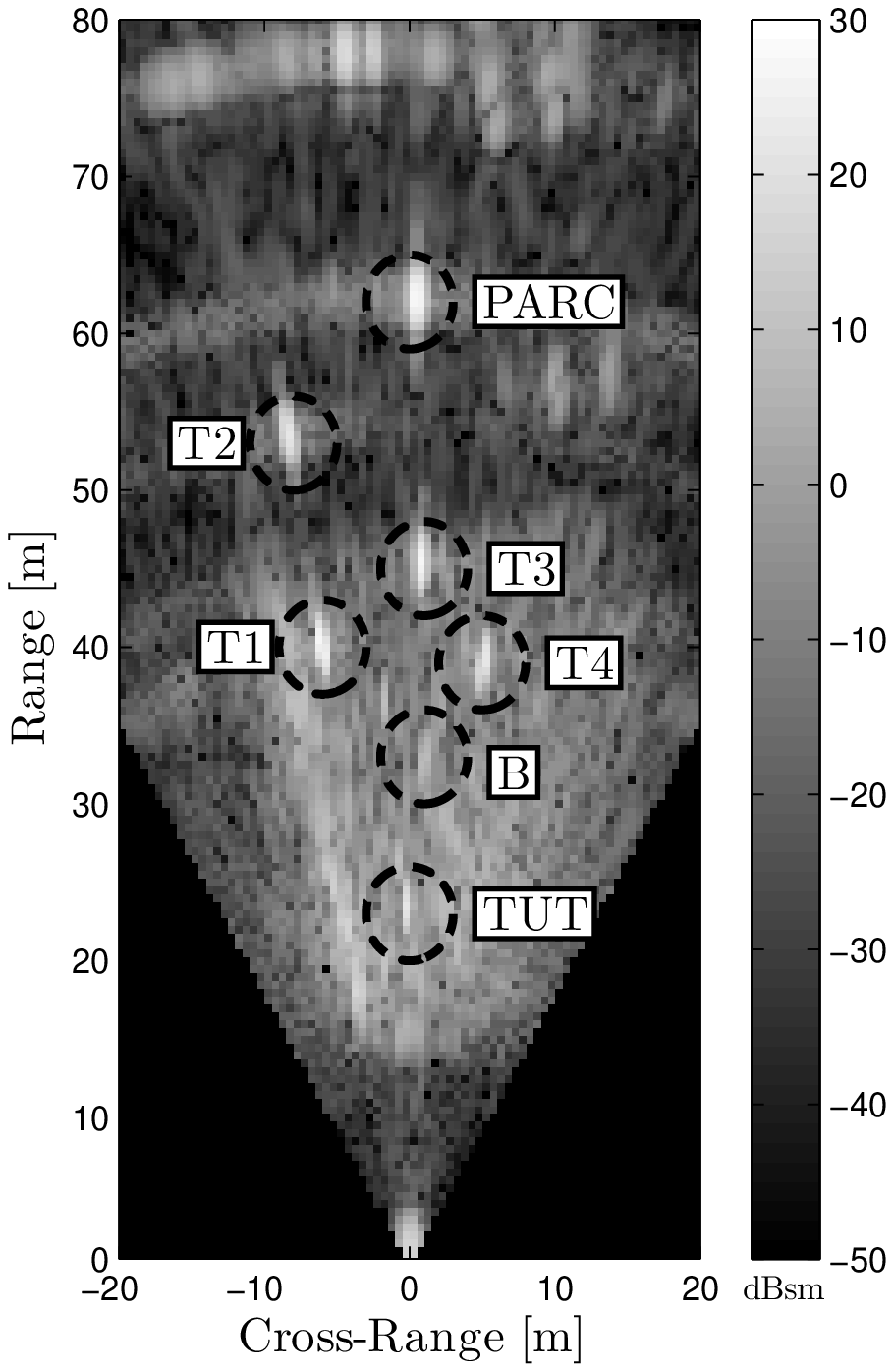}}\quad
\caption{Measured $|S_{HH}|$, $|S_{HV}|$, $|S_{VH}|$ and $|S_{VV}|$ channels for the TCR. \label{f:transpol_gbsar_meas_tcr}}
\end{figure}

The measured results for the case of a conventional TCR as trihedral under test (TUT) for co-polar ($|S_{HH}|$, $|S_{VV}|$) and cross-polar ($|S_{HV}|$, $|S_{VH}|$) channels are plotted in \fig{f:transpol_gbsar_meas_tcr}, considering a maximum range of 80 m, that is, zooming in the area of interest. It is seen that the TUT presents a high backscattered level in the co-polar channels, while its level is reduced in the cross-polar ones, as expected for a conventional TCR. This is confirmed as well with the backscattered signal of the reference TCRs (T1-T4). Note that the Bruderhedral (B) is only present in the cross-polar channels, as expected from a cross-polarising device. This is not the case of the PARC, which is clearly identified in all polarisations, due to its 45$^\circ$ tilt. It is worth noting that the amplification in the near range of our measurement scenario, mainly due to the proximity to the GB-SAR system, increases the floor level in the surroundings of the TUT, while slightly masking its polarimetric signature.

\begin{figure}[!h]
\centering
\subfigure[][TTCR $|S_{HH}|$]{\includegraphics[height=7cm]{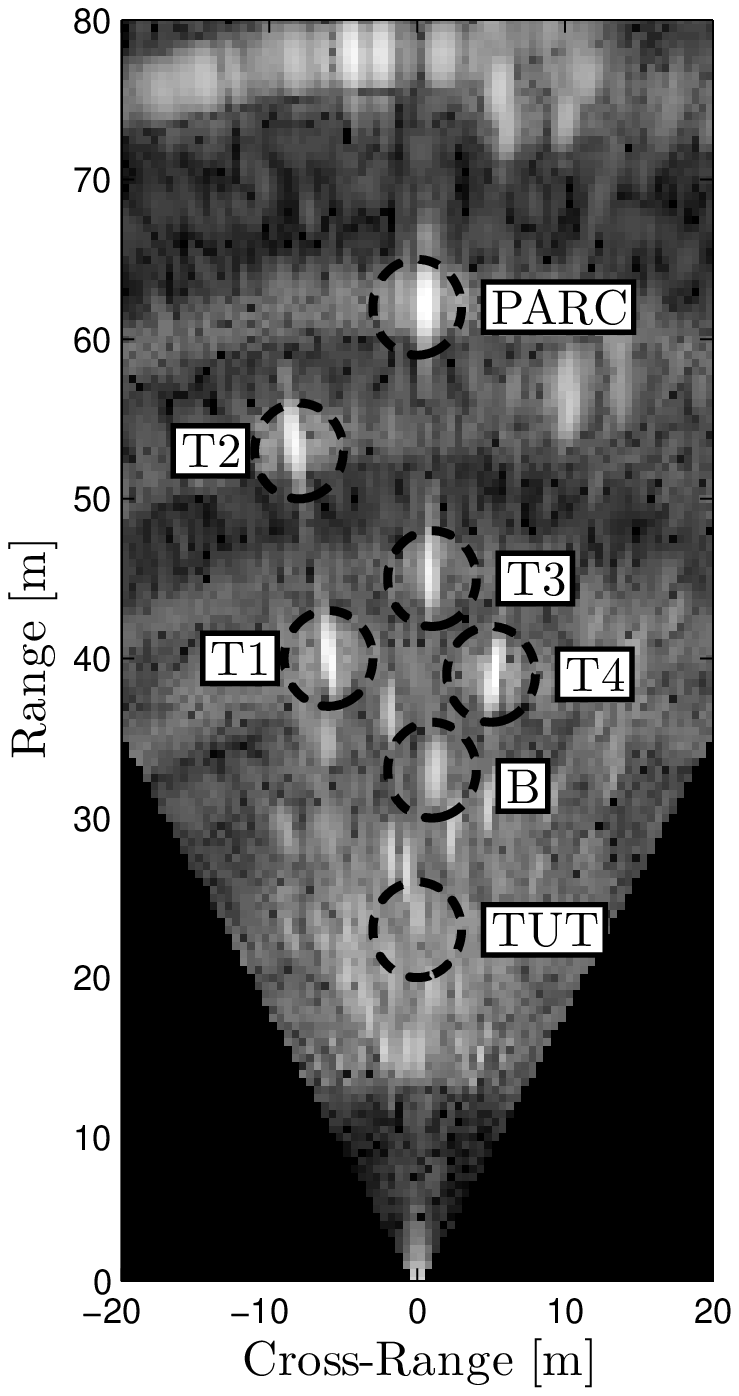}}\quad
\subfigure[][TTCR $|S_{HV}|$]{\includegraphics[height=7cm]{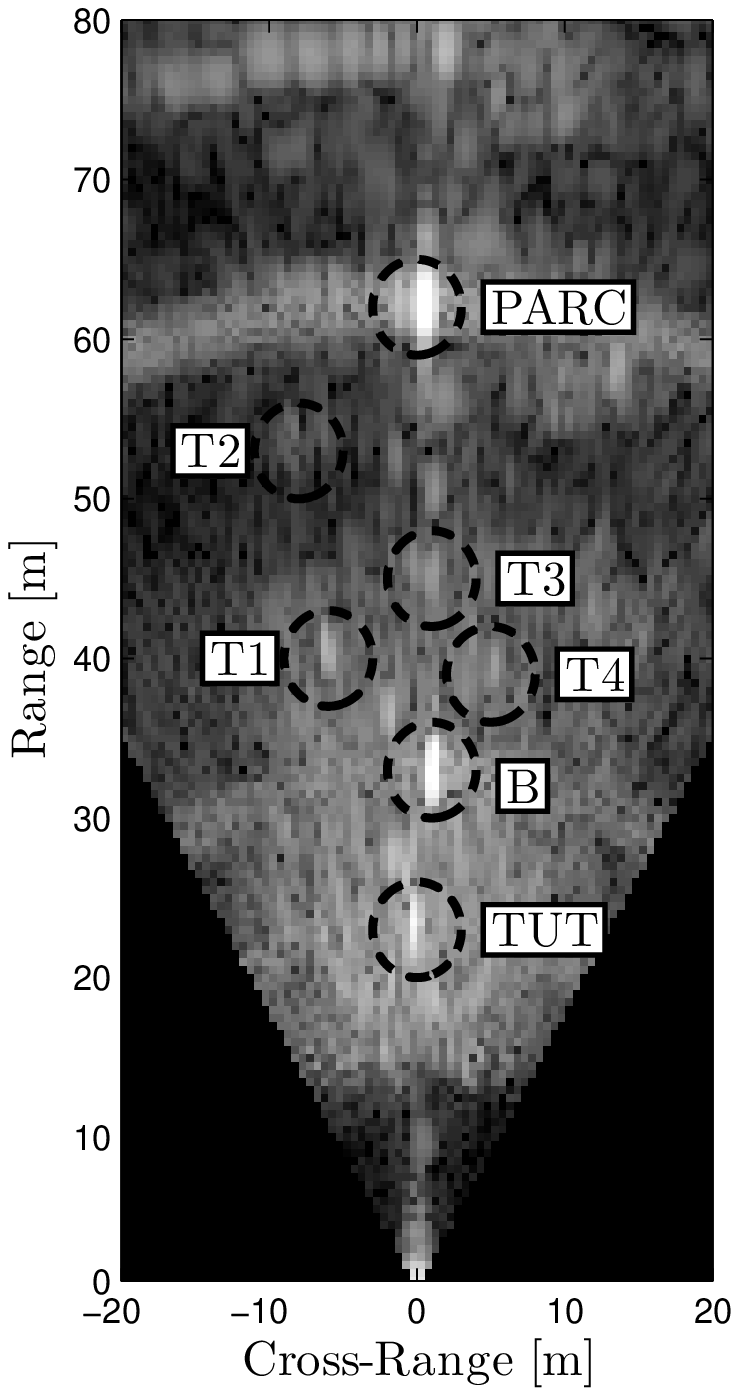}}\quad
\subfigure[][TTCR $|S_{VH}|$]{\includegraphics[height=7cm]{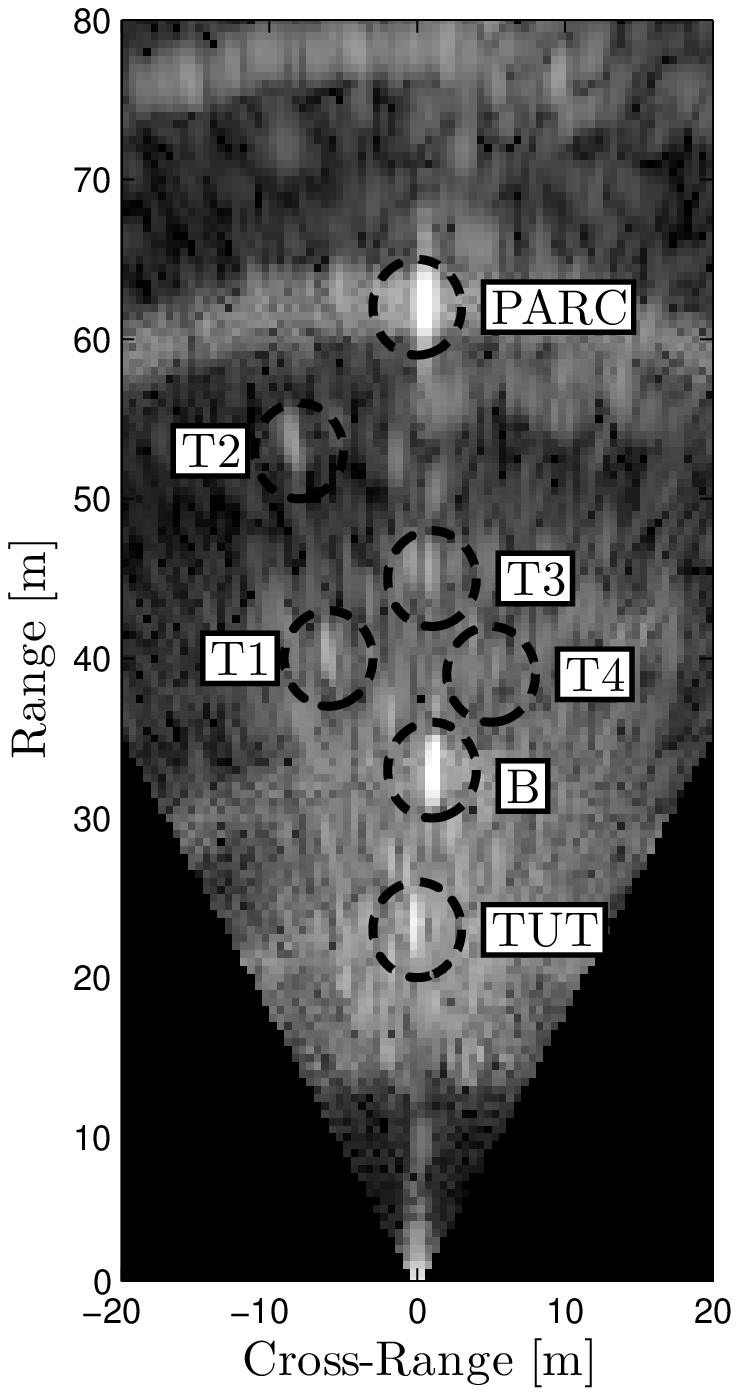}}\quad
\subfigure[][TTCR $|S_{VV}|$]{\includegraphics[height=7cm]{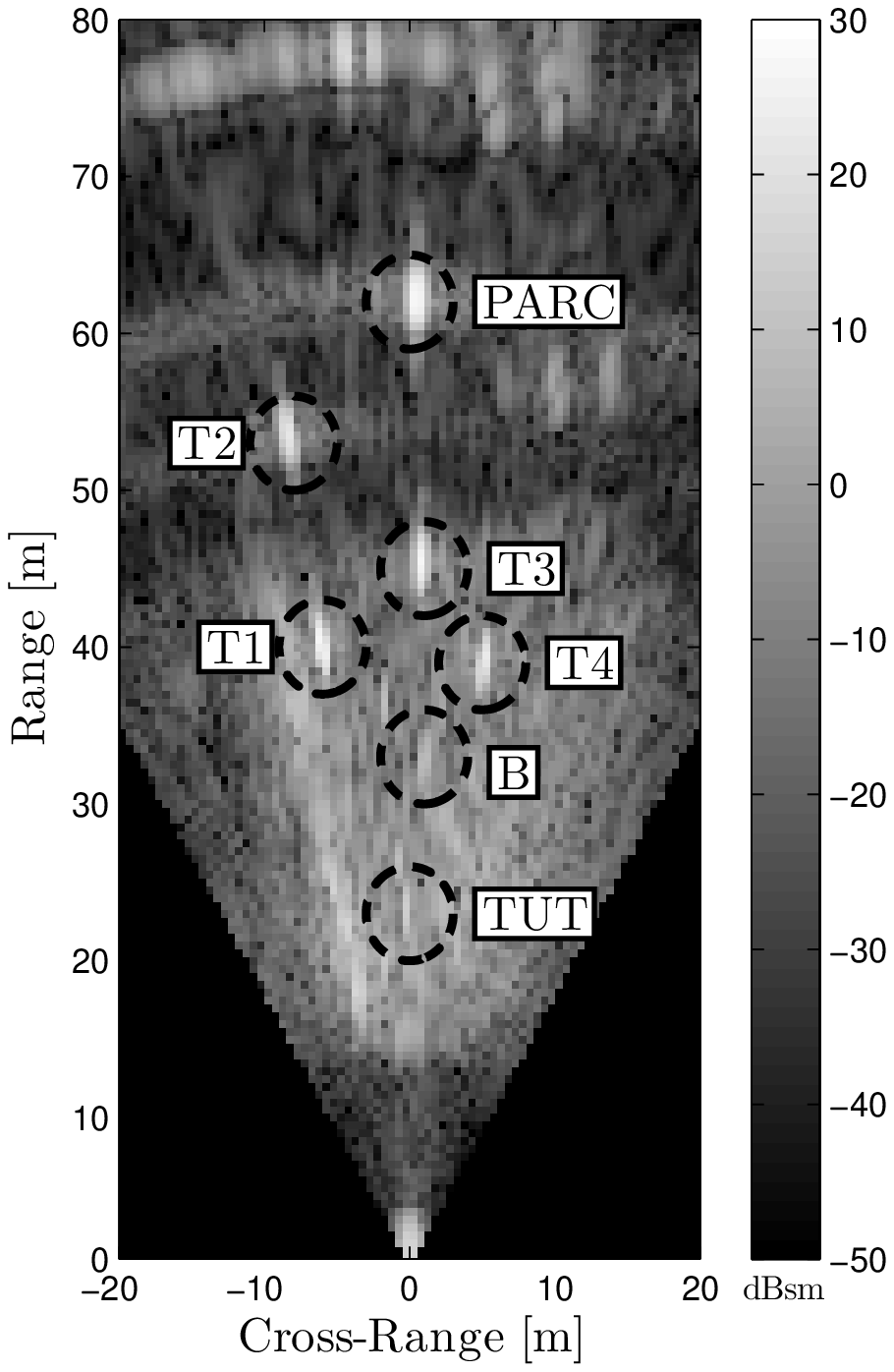}}\quad
\caption{Measured $|S_{HH}|$, $|S_{HV}|$, $|S_{VH}|$ and $|S_{VV}|$ channels for the TTCR. \label{f:transpol_gbsar_meas_ttcr}}
\end{figure}

The same measurements have been carried out for the case of a transpolarising-TCR (TTCR) as TUT. The magnitude of the co-polar ($|S_{HH}|$, $|S_{VV}|$) and cross-polar ($|S_{HV}|$, $|S_{VH}|$) channels are plotted in \fig{f:transpol_gbsar_meas_ttcr}. The reference TCRs are clearly identified in the co-polar results, as expected. But it can be seen that the TTCR presents a high cross-polar response, as well as the Bruderhedral and the PARC system, contrary to the case of a standard TCR as TUT. Moreover, although not completely vanished, the TTCR presents a low backscattered in the co-polar results, comparable to the cross-polar level found at the reference TCRs positions. 

\begin{figure}[!h]
\centering
\subfigure[][Cross-range cut for TCR]{\includegraphics[width=8.1cm]{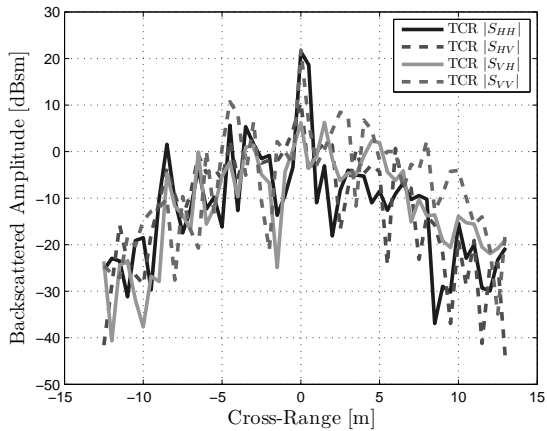}}\qquad
\subfigure[][Cross-range cut for TTCR]{\includegraphics[width=8.1cm]{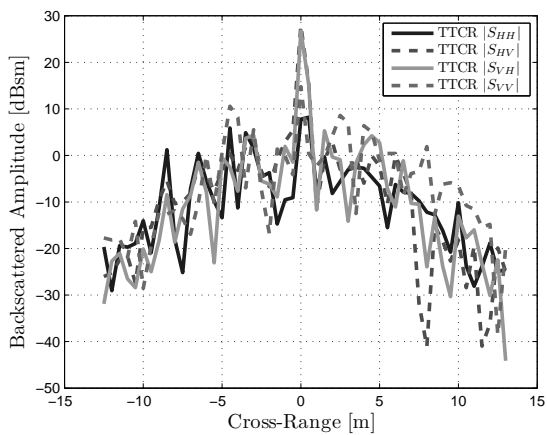}}\qquad
\caption{Measured cross-range cuts for the TCR and TTCR at the TUT position. \label{f:transpol_gbsar_meas_xrange_cuts}}
\end{figure}

The measured backscattered cross-range cuts for the four polarimetric channels at the TUT position are plotted in \fig{f:transpol_gbsar_meas_xrange_cuts}. The maxima backscattered results are well identified in the centre of the cross-range cuts. The TCR has high co-polar and low cross-polar responses, whereas the TTCR produces a reverse result to that of a TCR, and hence, the cross-polar enhancement of the TTCR is proved. The TTCR presents a cross-polarisation ratio of 12 dB at 9.65 GHz, outperforming the result obtained in the anechoic chamber.

The scattering matrices $($\ref{eq:transpol_rcs_tcr}$)-($\ref{eq:transpol_rcs_ttcr}$)$ have been also extracted from the GB-SAR measurements at 9.65 GHz. Despite some small disagreements, the scattering matrices are similar to those extracted from the anechoic chamber measurements $($\ref{eq:transpol_rcs_tcr_ttcr_anechoic}$)$. The TTCR presents an almost anti-diagonal matrix with a phase difference smaller than 20$^\circ$ between cross-polar components. Therefore, the GB-SAR measurement results prove the feasibility of the proposed TTCR as PolSAR calibrator.

\begin{equation}
\begin{array}{c}
\emph{\textbf{S}}_{TCR}^{meas} = \emph{A}\left[ \begin{array}{c c} 0.96_{\angle\;10^\circ} & 0.27_{\angle\;81^\circ}\\ 0.16_{\angle\;40^\circ} & 1.00_{\angle\;0^\circ} \end{array} \right]\\
\\
\emph{\textbf{S}}_{TTCR}^{meas} = \emph{A}\left[ \begin{array}{c c} 0.09_{\angle\;-117^\circ} & 1.00_{\angle\;0^\circ}\\ 0.88_{\angle\;-19^\circ} & 0.22_{\angle\;135^\circ} \end{array} \right]
\end{array}
\label{eq:transpol_rcs_tcr_ttcr_gbsar}
\end{equation}

\section{Conclusion}
In this letter, the performance of a transpolarising-TCR (TTCR) for PolSAR calibration purposes has been assessed with a GB-SAR system operating at 9.65 GHz (X-band). The TTCR presents a cross-polarisation ratio of 12 dB, and a phase difference of 19$^\circ$ between cross-polar components. These results could be improved by slightly retuning the transpolarising surface design in order to enhance the transpolarisation performance of the TTCR at 9.65 GHz. In addition, these measurements are affected by the amplification present in the near-range of the GB-SAR system, which increases the noise level; this fact may be overcome by using a bigger TTCR placed in a farther and cleaner region in the measurement scenario. The TTCR design is characterised by its low profile and light weight transpolarising surface, which can be easily fabricated by applying standard photo-etching techniques, while overcoming some practical fabrication aspects of previous TTCR designs (gridded and corrugated TCRs). Finally, the proposed TTCR design could also be applied to other frequency bands (e.g., L or C bands), by scaling, in terms of wavelength, the dimensions of the elements (square patches with slots) of the transpolarising surface to the corresponding frequency of operation.

%
%

\ifCLASSOPTIONcaptionsoff
  \newpage
\fi

\end{document}